\documentclass{PoS}

\title{Diffractive production of isolated photons at HERA}

\ShortTitle{Diffractive photons}

\author{\speaker{Peter Bussey}\\
        School of Physics and Astronomy\\
        University of Glasgow\\
        Glasgow, United Kingdom, G12 8QQ.   
        E-mail: \email{peter.bussey@glasgow.ac.uk\\  \\ }
        For the ZEUS Collaboration.}

\newcommand{\text}{\rm}
\newcommand{\xspace}{ }
\newcommand{\eVdist}{\kern-0.06667em}
\newcommand{\gev}{{\,\rm{Ge}\eVdist\rm{V\/}}}

\newcommand{\PYTHIA}{\textsc{Pythia}}
\newcommand{\HERWIG}{\textsc{Herwig}}
\newcommand{\RAPGAP}{\textsc{Rapgap}}

\newcommand{\ET}{\ensuremath{E_{T}}\xspace}
\newcommand{\ETjet}{\ensuremath{E_{T}^{\mathrm{jet}}}\xspace}
\newcommand{\Ejet}{\ensuremath{E^{\mathrm{jet}}}\xspace}
\newcommand{\pzjet}{\ensuremath{p_Z^{\mathrm{jet}}}\xspace}

\newcommand{\Egam}{\ensuremath{E^{\gamma}}\xspace}
\newcommand{\pzgam}{\ensuremath{p_Z^{\gamma}}\xspace}
\newcommand{\ETgam}{\ensuremath{E_{T}^{\gamma}}\xspace}
\newcommand{\etagam}{\ensuremath{\eta^{\gamma}}\xspace}
\newcommand{\etajet}{\ensuremath{\eta^{\mathrm{jet}}}\xspace}
\newcommand{\etamax}{\ensuremath{\eta^{\mathrm{max}}}\xspace}

\newcommand{\xgamm}{\ensuremath{x_{\gamma}^{\mathrm{meas}}}\xspace}
\newcommand{\pom}{\mbox{\scriptsize\it I\hspace{-0.5ex}P}}
\newcommand{\xpom}{\ensuremath{x_{\pom}}\xspace}
\newcommand{\zpom}{\ensuremath{z_{\pom}^{\mathrm{meas}}}\xspace}

\newcommand{\etal}{{\it et al.}}

\abstract{The ZEUS detector at HERA has been used to measure  the
photoproduction of isolated photons  
in diffractive events.   Cross sections are
evaluated in the photon transverse-energy and pseudorapidity ranges
\mbox{$5 < E_T^{\gamma} < 15\,\gev$} and \mbox{$-0.7 < \eta^{\gamma} <
0.9$,} inclusively and with a jet with transverse-energy an
  pseudorapidity in the ranges $4 <
\ETjet < 35\,\gev$ and $-1.5 < \etajet < 1.8$, for an integrated
luminosity of 374 $\mathrm{pb}^{-1}$. Further kinematic variables
studied include the fractions of the incoming photon energy and of the
colourless exchange (``Pomeron'') energy that are imparted to a photon-jet
final state.  Comparison is made to 
predictions from the \RAPGAP\ Monte Carlo simulation.  
}

\FullConference{XXIII International Workshop on Deep-Inelastic Scattering,\\
		27 April - May 1 2015\\
		Dallas, Texas}

\begin{document}
\section{Introduction}

%%%%%%%%%%%%%%%%%%%%%%%%%%%
%Figure 1

\begin{figure}[b]
\begin{center}
\includegraphics[width=0.45\linewidth]{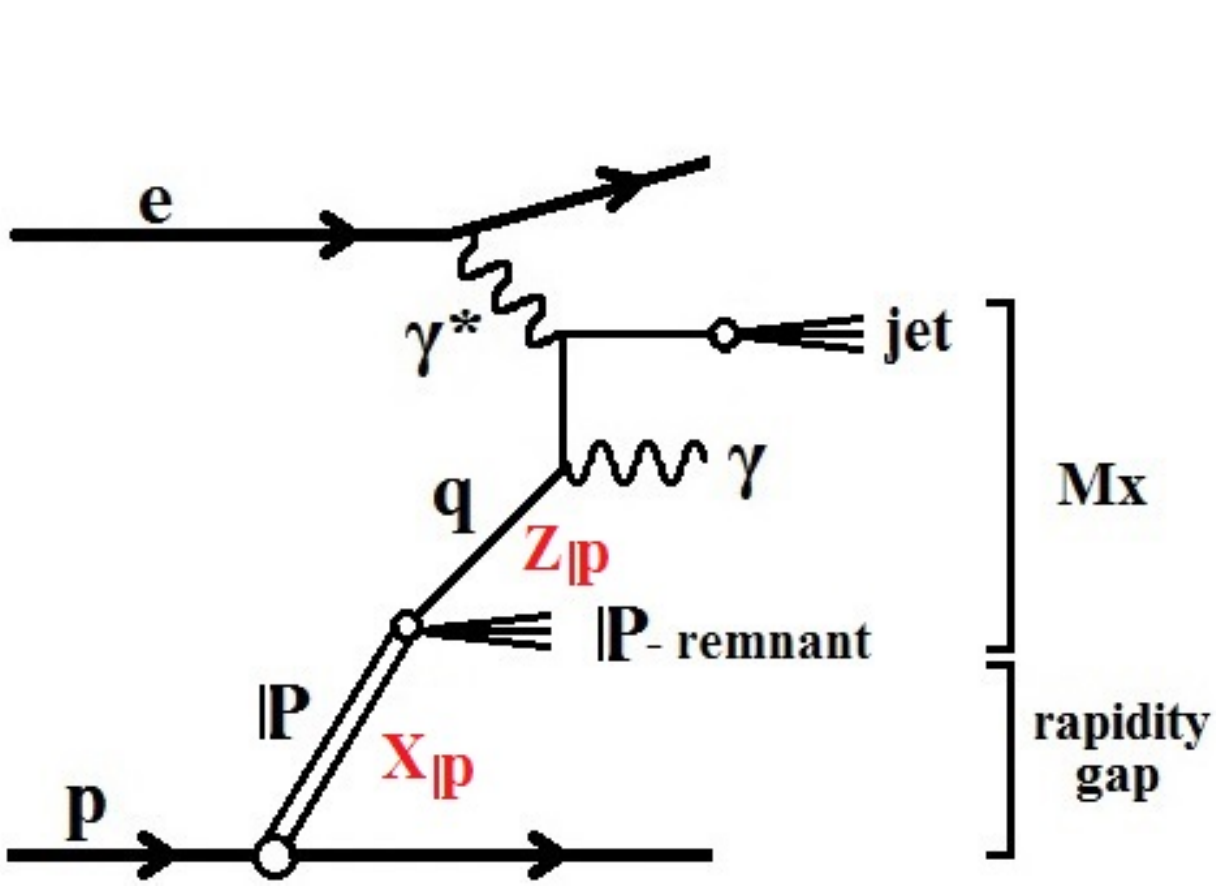}
\includegraphics[width=0.45\linewidth]{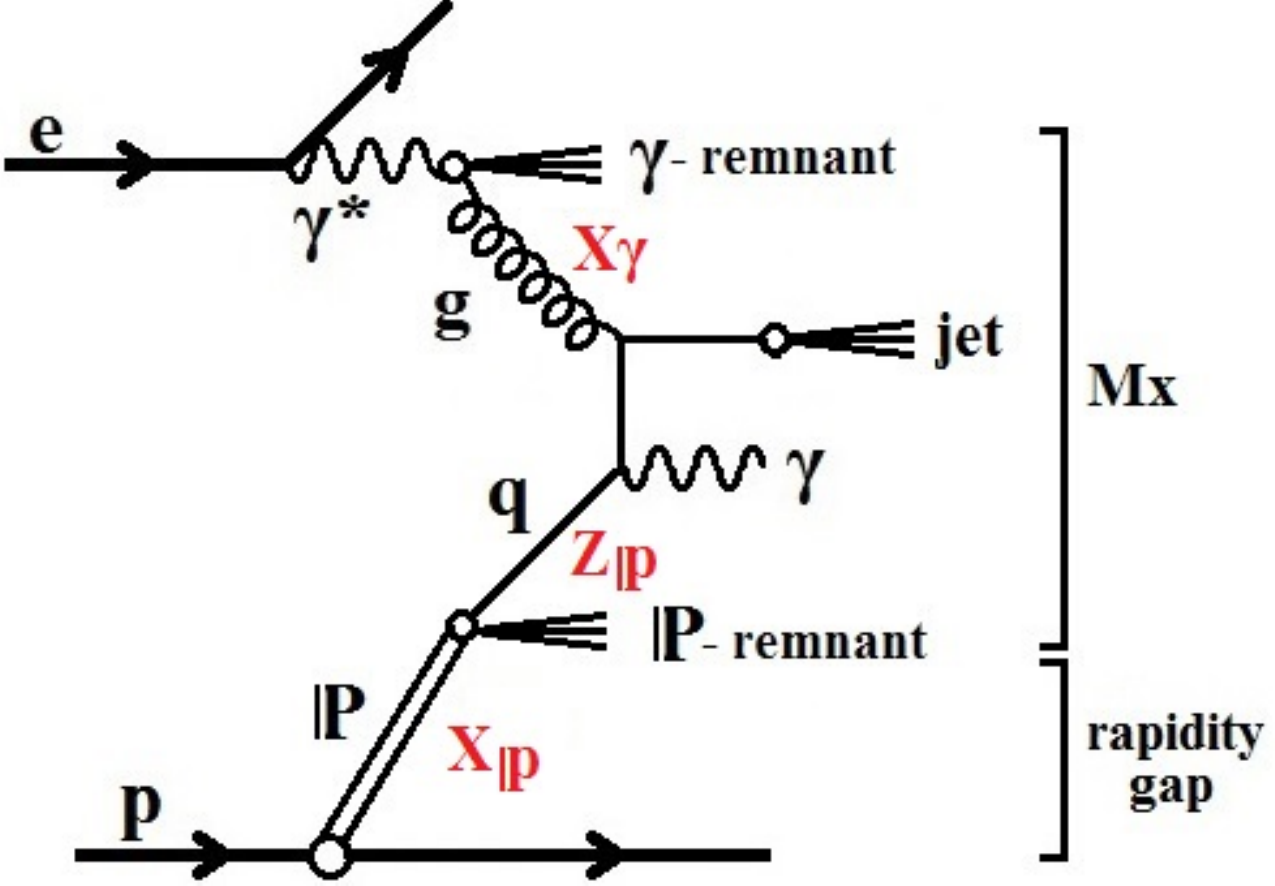}
~\\[-35mm]
\end{center}
\vspace*{-35mm}\hspace*{0.5cm}
(a)\hspace{7.5cm}
(b)\hspace*{7.5cm}\\[25mm]
\vspace*{0mm}
\caption{Examples of the diffractive production of a prompt photon
and a jet in $ep$ scattering from (a) direct (b) resolved photons.} 
\label{fig1}
\end{figure}

  In diffractive hadronic particle interactions, a colour-neutral object is
exchanged, frequently referred to as the ``pomeron''. Diffractive
scattering off protons may be initiated by a second incoming hadron,
or even by a real or virtual photon.  At the HERA $ep$ collider,
diffractive processes have been studied both in photoproduction and in
deep inelastic scattering, the photoproduction processes consisting of
those in which the exchanged virtual photon is quasi-real. The
diffractive process is characterised by a forward nucleon, 
followed by a ``gap'' in rapidity in which little or no energetic scattering is
found until the central region of the process where the hard final
state is detected and measured.

This talk presents measurements in which a hard isolated ``prompt'' photon is detected in
the central region of the ZEUS detector and may be accompanied by a
jet.  Such processes, while rare, are interesting for several
reasons.  An outgoing photon must be radiated from a charged partonic
object, namely a quark, and therefore demonstrates the presence of a
quark content in the pomeron or of higher-order processes in which
both the pomeron and incident photon couple to quarks.  In general, 
they allow QCD-based models to be tested in order to improve our understanding
of a type of process which is important at high energies.

The observed processes also include ``fragmentation processes'' in which a
photon is radiated within a jet. Such processes are suppressed by
requiring that the outgoing photon be isolated.
The present measurements follow an earlier study by
H1~\cite{plb:672:219} of inclusive diffractive high energy prompt photons as a
function of transverse momentum.  Analyses of isolated photons in
non-diffractive photoproduction have also been made by the ZEUS and H1
collaborations~\cite{pl:b730:293,pl:b413:201}. 
% \cite{pl:b413:201,pl:b472:175,pl:b511:19,epj:c49:511,epj:c38:437,epj:c66:17}, 
%as well as in deep inelastic scattering~\cite{pl:b595:86}.
% \cite{pl:b595:86,epj:c54:371,pl:b687:16,pl:b715:88}. 

\section{Kinematic quantities}
\label{kin-q}

In ``direct'' photoproduction processes, the entire incoming photon is
absorbed by an outgoing quark from the incoming proton, while in
``resolved'' processes, the photon's hadronic structure provides a
quark or gluon that interacts with a parton from the proton.  These
two classes of process (fig.~\ref{fig1}), which are unambiguously defined only at lowest
order, may be partially distinguished in events containing a high-\ET\
photon and a jet by means of the quantity
$$
\xgamm =  \frac{\Egam+\Ejet-\pzgam-\pzjet}
{E^{\text{all}}- p_Z^{\text{all}}}, 
$$ which measures the fraction of the incoming photon energy that is
given to the photon and the jet. Similarly, direct and resolved
pomeron processes may be defined. The quantities \Egam\ and \Ejet
denote the energies of the photon and the jet, respectively, $p_Z$
denotes the corresponding longitudinal momenta, and the suffix ``all''
refers to all the final-state particles or detector-measured objects
in an event. The presence of direct processes generates a prominent
peak in the cross section at high \xgamm values.

When the  proton, with energy $E_p$, radiates an object such as a pomeron, which interacts
with all or part of an incoming photon, the fraction of the proton energy
taken by the radiated pomeron is given to a good approximation by:
$$
 \xpom =(E^{\text{all}}+ p_Z^{\text{all}})/2E_p.
$$

The fraction of the pomeron energy that takes part in the hard
interaction that generates the outgoing photon and jet is given
to a good approximation by: 
$$
\zpom =  \frac{\Egam+\Ejet+\pzgam+\pzjet}{E^{\text{all}}+ p_Z^{\text{all}}}.
$$

\section{Experimental method}
\label{sec-exp}
The measurements are based on two data samples corresponding an
integrated luminosities of 91 and $374\,\mathrm{pb}^{-1}$, taken
during the years 1998-2000 and 2004-2007 respectively with the ZEUS
detector at HERA, and referred to as HERA-I and HERA-II samples respectively.  During these periods, HERA ran with an electron or
positron beam energy of 27.5\gev\ and a proton beam energy of $E_p
=920$\gev. The samples include  $e^+p$ and $e^-p$
data\footnote{Hereafter ``electron'' refers to both electrons and positrons unless otherwise stated.}.

A detailed description of the ZEUS detector is given 
elsewhere~\cite{zeus:1993:bluebook}. Charged particles were measured in
the central tracking detector and a silicon micro
vertex detector which operated in a magnetic
field of $1.43$~T provided by a thin superconducting solenoid.  The
barrel electromagnetic calorimeter (BEMC) cells had a pointing
geometry aimed at the nominal interaction point. 
Its fine granularity allows the use of shower-shape
distributions to distinguish isolated photons from the products of
neutral meson decays such as $\pi^0 \rightarrow \gamma\gamma$.

%%%%%%%%%%%%%%%%%%%%%%%%%%%%%%%%%%%%%%%%%%%%%%%%%%%%%%%%%%%%%%%%

Monte Carlo (MC) event samples were employed to evaluate the detector
acceptance and event-reconstruction efficiency, and to provide signal
and background distributions.  The program \RAPGAP\ 3.2~\cite{rapgap}
was used to generate the diffractive process $pe \to pe\gamma X$
for direct and resolved incoming virtual photon exchange, where $X$ denotes
any other particles.  The
diffractive proton pdf set H1-B (2006) was used, and for the resolved
photon the pdf set SaSG 1D LO.  The program \PYTHIA\
6.416~\cite{jhep:0605:026} was used to generate direct and resolved
prompt-photon photoproduction processes for background calculations,
and also $2\to2$ parton-parton scattering processes not involving
photons (``dijet events''), making use of the CTEQ4 and GRV proton and
photon parton densities.  The program was run using the default
parameters with minor modifications. Non-diffractive photoproduction
event samples were also generated using the \HERWIG\ 6.510
program~\cite{jhep:0101:010}, again with minor modifications to the
default parameters.  The \PYTHIA\ and \HERWIG\ programs differ
significantly in their treatment of parton showers, and in the use of
a string-based hadronisation scheme in \PYTHIA\ but a cluster-based
scheme in \HERWIG.

The measured photons are accompanied by backgrounds from neutral
mesons in hadronic jets, in particular $\pi^0$ and $\eta$, where the
meson decay products create an energy cluster in the BCAL that passes
the selection criteria for a photon.  Although not generated
diffractively, the dijet event samples enabled such background events
to be distinguished in the analysis..

%%%%%%%%%%%%%%%%%%%%%%%%%%%%%%%%%%%%%%%%%%% Event selection

The basic event selection and reconstruction were performed as
previously~\cite{pl:b730:293}.  A three-level trigger system was used
to select events online.  
The event analysis made use of energy-flow objects
(EFO's), constructed
from clusters of calorimeter cells, associated with
tracks when appropriate, and also unassociated tracks. Photon candidates were identified as
EFO's with no associated track, and with at least $90\%$ of the
reconstructed energy measured in the BEMC. 

Jets were reconstructed using all the EFO's in the
event, including photon candidates, by means of the $k_T$ clustering
algorithm in the $E$-scheme in the longitudinally
invariant inclusive mode with the radius parameter
set to 1.0. 
To reduce the fragmentation contribution and the background from the
decay of neutral mesons within jets, the photon candidate was required
to be isolated from other hadronic
activity.  This was imposed by requiring that the photon-candidate EFO
had at least 90\% of the total energy of the reconstructed jet of
which it formed a part.

Events were finally selected with the following kinematic conditions:
Each event was required to contain a photon candidate with a 
reconstructed transverse energy, $E_T^{\gamma}$, in the range
\mbox{$5 <E_T^{\gamma}<15\,\mathrm{\gev\ }$} and with pseudorapidity,
$\eta^{\gamma}$, in the range $-0.7 < \eta^{\gamma} < 0.9$;  
The hadronic jet, when used,  was required to have  
$E_T^{\mathrm{jet}}$ between 4 and 35\gev\ and to lie within the
pseudorapidity, $\eta^{\mathrm{jet}}$, range $-1.5
<\eta^{\mathrm{jet}} < 1.8$;
To select diffractive events further conditions were applied,
the first of which was that the maximum pseudorapdity for EFO's  with energy
above 0.4 GeV satisfied $\eta_{\mathrm{max}} < 2.5$. 
A second diffractive condition was that $\xpom < 0.03$. 
For the HERA-I data sample, the energy in the Forward Plug 
Calorimeter was required to be less than 1 GeV.  This calorimeter was
not present in the HERA-II running.

The HERA-II data sample was used in the main analysis described here.
The HERA-I data were analysed similarly, with the addition of the
selection on the Forward Plug Calorimeter, and were used to provide
a correction as described below. A further background arises from Deeply Virtual Compton Scattering events, which were  removed from the sample.

\begin{figure}[t]
\begin{center}
\includegraphics[width=0.45\linewidth]{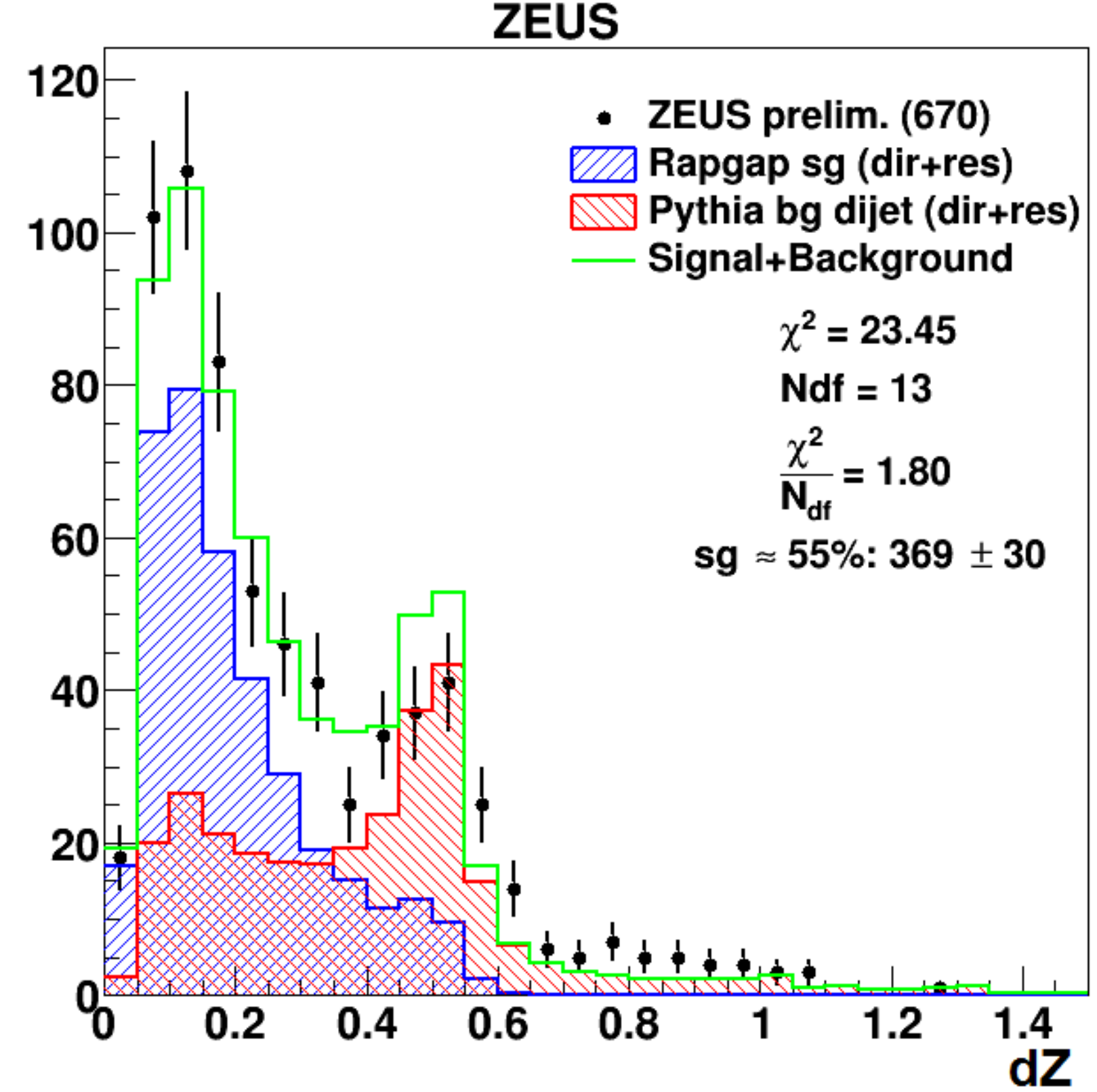}
\hspace{5mm}
\includegraphics[width=0.475\linewidth]{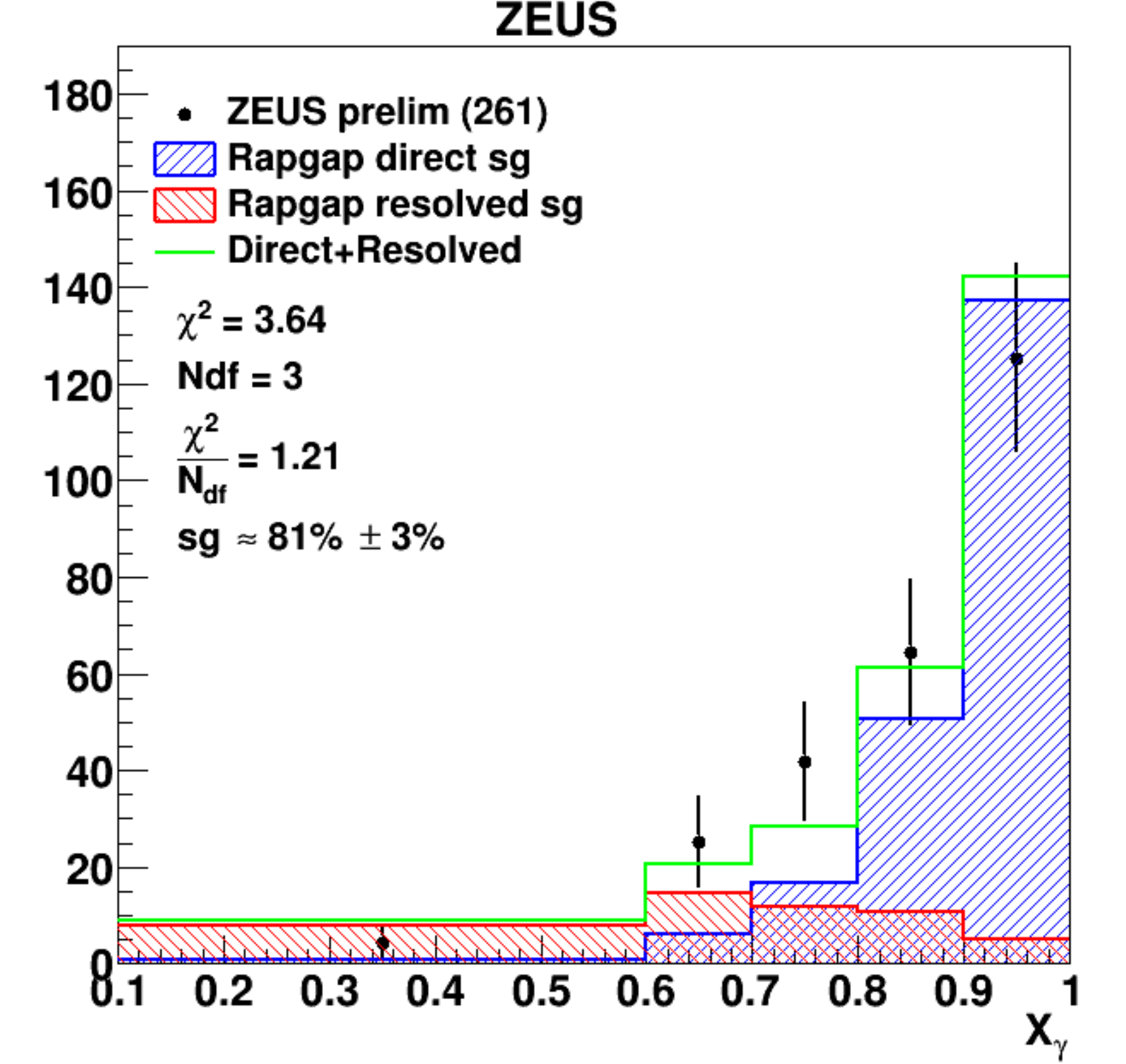}
\end{center}
\vspace*{-35mm}\hspace*{5.5cm}
(a)\hspace{2.5cm}
(b)\hspace*{7.5cm}\\[25mm]
\vspace*{0mm}
\caption{ 
(a) Distribution of $\langle \delta Z \rangle$ for 
  selected photon candidates. 
(b) Distribution of photon events, after experimental selections, as a function of  \xgamm per unit interval of 0.1 in \xgamm, compared to  
a mixture of reweighted \RAPGAP-generated direct and resolved events using the
model described in the text.  }
\label{fig:showers}
\end{figure}

The selected samples include background events arising from neutral
meson decays.  The photon signal was extracted statistically following
the approach taken in previous ZEUS analyses, and used the
energy-weighted width measured in the $Z$ direction, $\langle
dZ\rangle$, of the BEMC energy cluster that comprised the photon
candidate.  For each measured cross-section bin, the number of
isolated-photon events in the data was determined by a
Maximum-Likelihood fit to the $\langle \delta Z \rangle$ distribution
in the range $0.05<\langle \delta Z \rangle < 0.8$, varying the
relative fractions of the signal and background components as
represented by histogram templates obtained from the MC
(fig.~\ref{fig:showers}(a)).

After applying the selections described above, the distribution of
the events in \xgamm\ is as shown in fig.~\ref{fig:showers}(b).   A 80:20
mixture of direct:resolved \RAPGAP\ events gives a reasonable description
of the data and was used subsequently. 
To evaluate the acceptances, allowance must be made for the different
acceptances for the direct and the resolved processes.  A weighting 
factor as a function of \etamax\ was applied to give agreement between
the \RAPGAP\ prediction and the data; however this has only a small effect on
the acceptance calculation.    A bin-by-bin correction method was then used
to determine the production cross section in a given variable.

Two further corrections were applied to the HERA-II data, which were used
to provide the results presented below.  The total cross sections for the
diffractive processes were evaluated, and the HERA-II
data were rescaled to have the same total cross section as measured using
the \mbox{HERA-I} data.  This compensated for a potential loss of events
in the HERA-II data due to scattering of forward-generated particles off
a beamline magnet that  occupied most of the central aperture
of the forward ZEUS calorimeter in the HERA-II running, but which was
absent during the HERA-I running.  The \mbox{HERA-I} Forward Plug Calorimeter
removed additional non-diffractive events in this angular region.

A second correction was needed to remove non-diffractive events that
succeeded in passing the diffractive event selections described above.
It was evaluated using \PYTHIA\ and \HERWIG\
photoproduction event samples, which were normalised to the data
sample after extracting the photon signal without applying the
diffractive event selections. The diffractive event
selections were then applied to the Monte Carlo samples to estimate
the contribution of the photoproduction to the observed data.  
The mean of the two corrections was applied.

\section{Results}

\label{sec:results}

Differential cross sections were calculated for the diffractive
production of an isolated photon, inclusive and with at least one
accompanying jet.  The kinematic region was defined by $Q^2< 1 \,
\mathrm{\gev}^2$, $0.2 < y < 0.7,$ $-0.7< \etagam < 0.9$, $5 < \ETgam<
15\,\mathrm{\gev} $, $4 < \ETjet < 35$\gev and $-1.5 <\etajet<
1.8$. The diffractive condition consisted in requirements that $\xpom
<0.03$ and $\etamax<2.5$.  These quantities were evaluated at the
hadron level in the laboratory frame, matching the cuts used at the
detector level, and the jets were formed according to the $k_T$
clustering algorithm with the radius parameter set to 1.0.  Photon
isolation was imposed such that at least $90 \%$ of the energy of the
jet-like object containing the photon originated from the photon.

\begin{figure}[t]
\vfill
\begin{center}
\includegraphics[width=.45\linewidth]{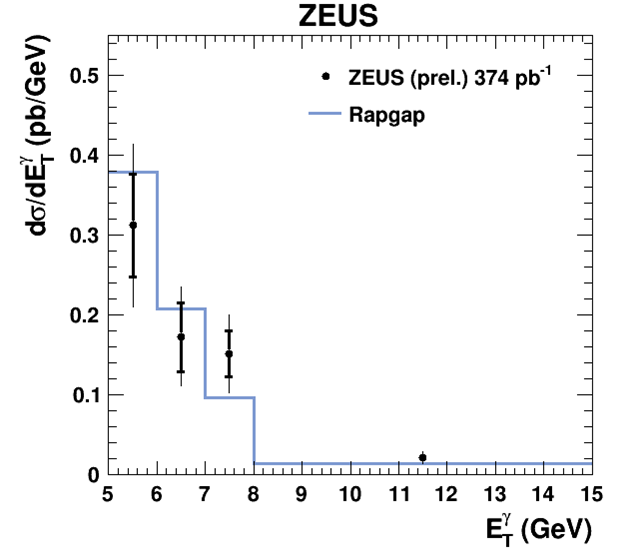}
\includegraphics[width=.45\linewidth]{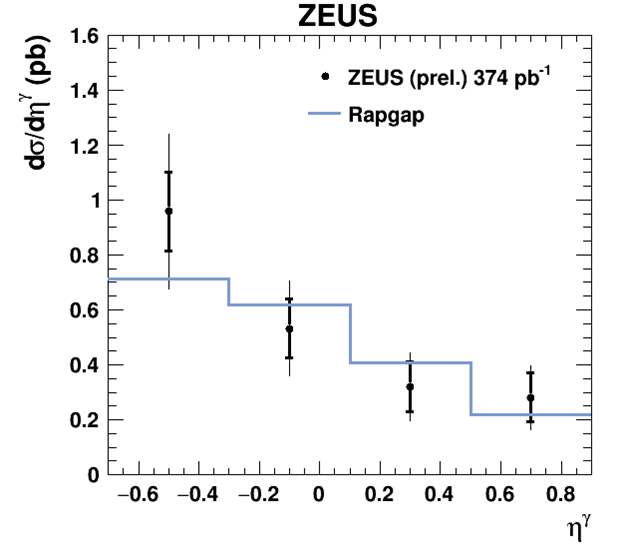}
\end{center}
\vspace*{-60mm}\hspace*{2.0cm}
(a)\hspace{6.5cm}
(b)\hspace*{7.5cm}\\[50mm]
\vspace*{0mm}
\begin{center}
\includegraphics[width=.45\linewidth]{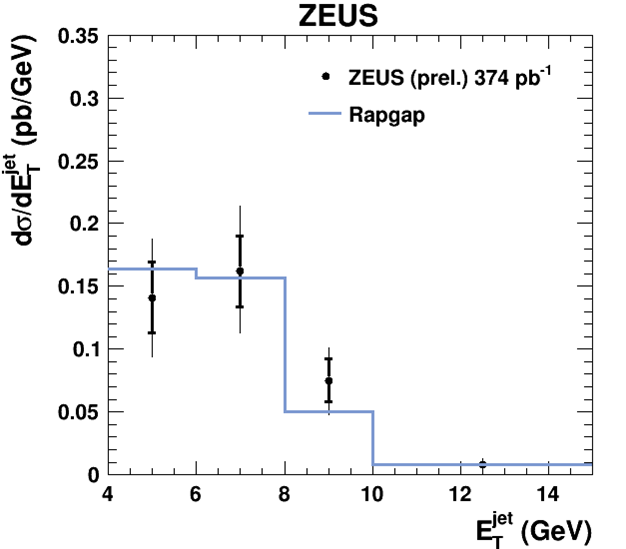}
\includegraphics[width=.45\linewidth]{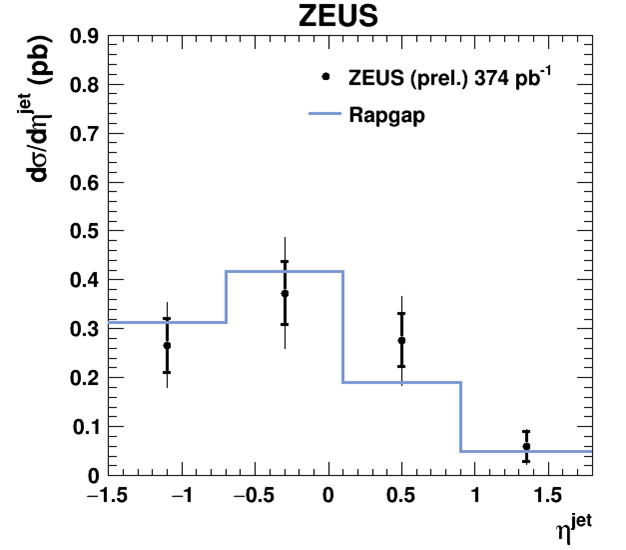}
\end{center}
\vspace*{-60mm}\hspace*{2.0cm}
(c)\hspace{6.5cm}
(d)\hspace*{6.5cm}\\[50mm]
\vspace*{0mm}
\begin{center}
\includegraphics[width=.45\linewidth]{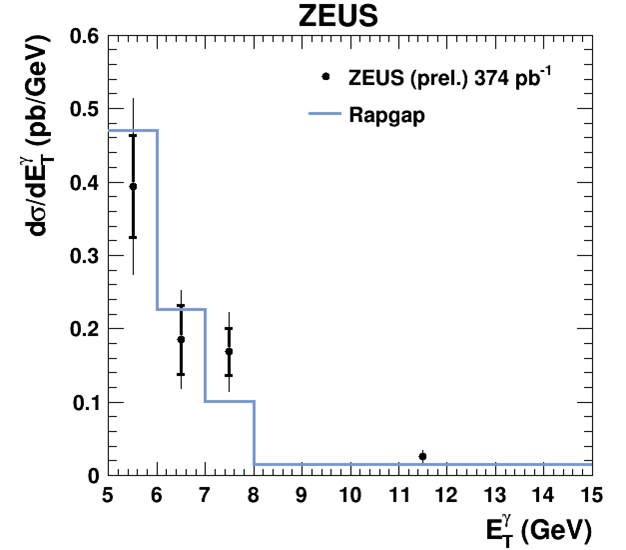}
\includegraphics[width=.45\linewidth]{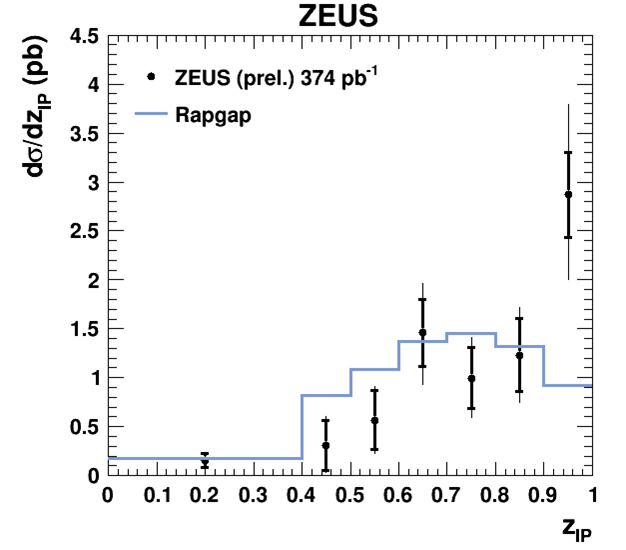}
\end{center}
\vspace*{-60mm}\hspace*{2.0cm}
(e)\hspace{11.0cm}
(f)\hspace*{7.5cm}\\[50mm]
\vspace*{0mm}
\caption{  Differential cross sections as functions of (a) \ETgam, (b) \etagam, 
(c) \ETjet, and (d) \etajet, 
for events containing an isolated photon accompanied by a jet,
(e) \ETgam  
for inclusive events containing an isolated photon, and 
(f) \zpom  
for events containing an isolated photon and a jet.
Comparisons are made to normalised predictions from \RAPGAP,
evaluated without reweighting.
}
\label{fig:res}
\end{figure}

Results are shown in fig.~\ref{fig:res}, together
with predictions from \RAPGAP, normalised to the data.
Cross sections in $\ETjet$ above 15\gev\ are omitted from 
fig.~\ref{fig:res}(c) owing to limited statistics, but this
kinematic region is included in the other cross-section measurements.
The results here are inclusive of proton excitation processes
which have been estimated to comprise approximately 16\% of the total
diffractive cross section.

    The most significant sources of systematic uncertainty arose
from the scaling of the HERA-II cross sections to the HERA-I data,
treated as a systematic uncertainty although its origin is statistical.
This amounts to an uncertainty of $~\pm20\%$.  The next most significant
source comes from the uncertainty in the non-diffractive correction
and was typically $\pm10\%$.    A further uncertainty can be estimated
by using the \HERWIG\ MC to model the signal and background events.
Less important uncertainties arose from the calibrations of the photon
and jet energy scales, the modelling of the resolved and direct ratios
in the acceptance calculation and the modelling of the background.
The effect of all of these was estimated by including an overall common 
scaling uncertainty of $\pm25\%$ on all the cross sections.  The
narrow uncertainty bars on the figures include this in combination with
the statistical uncertainties, which are indicated by a thick bar.

The cross section for photons with a jet is seen to be a high fraction of the
inclusive photon cross section.   For the quantity \zpom the data show a peak at the upper
end of the distribution which is not well described by \RAPGAP.
In all other cases, \RAPGAP\ gives a good description of the data.

%====================\input{Conclusions}

%%%%%%%%%%%%%%%%%%%%%%%%%%%%%%%%%%%%%%%%%%%%%%%%%Figure 5a


\begin{thebibliography}{99}
\bibitem{plb:672:219}
H1 Collaboration, V. Andreev \etal,
\newblock Phys.\ Lett.\ B 672 (2009) 219\relax
\relax
\bibitem{pl:b730:293}
ZEUS Collaboration, H. Abramowicz \etal,
 Phys.\ Lett.\ B~730~(2014)~293\relax 
\relax
%\bibitem{pr:d73:094007}
\bibitem{pl:b413:201}
ZEUS Collaboration, J.~Breitweg \etal,
\newblock Phys.\ Lett.{} B~413~(1997)~201\relax
\relax
\\ % \bibitem{pl:b472:175}
ZEUS Collaboration, J.~Breitweg \etal,
\newblock Phys.\ Lett.{} B~472~(2000)~175\relax
\relax
\\ % \bibitem{pl:b511:19}
ZEUS Collaboration, S.~Chekanov \etal,
\newblock Phys.\ Lett.{} B~511~(2001)~19\relax
\relax
\\ % \bibitem{epj:c49:511}
ZEUS Collaboration, S. Chekanov \etal,
\newblock Eur.\ Phys.\ J.{} C~49~(2007)~511\relax
\relax
\\ % \bibitem{epj:c38:437}
H1 Collaboration, A.~Aktas \etal,
\newblock Eur.\ Phys.\ J.{} C~38~(2004)~437\relax
\relax
%\bibitem{epj:c66:17}
%H1 Collaboration, F.D.~Aaron \etal, Eur.\ Phys.\ J.{} C~66~(2010)~17\relax
%\bibitem{epj:c54:371}
%H1 Collaboration, F.D.~Aaron \etal,
%\newblock Eur.\ Phys.\ J.{} C~54~(2008)~371\relax
%\relax
%\\ % \bibitem{pl:b595:86}
%ZEUS Collaboration, S.~Chekanov \etal,
%\newblock Phys.\ Lett.{} B~595~(2004)~86\relax
%\relax
%\\ % \bibitem{pl:b687:16}
%ZEUS Collaboration, S.~Chekanov \etal,
%\newblock Phys.\ Lett.{} B~687~(2010)~16\relax
%\relax
%\\ % \bibitem{pl:b715:88}
%ZEUS Collaboration, H. Abramowicz \etal,
%\newblock Phys.\ Lett.{} B~715~(2012)~88\relax
%\relax
\bibitem{zeus:1993:bluebook}
ZEUS Collaboration, U.~Holm~(ed.),
\newblock {\em The {ZEUS} Detector}.
\newblock Status Report (unpublished), DESY (1993),
\newblock available on
  \texttt{http://www-zeus.desy.de/bluebook/bluebook.html}\relax
\relax
\bibitem{rapgap}
H. Jung, \verb{http://www.desy.de/~jung/rapgap.html{
\bibitem{jhep:0605:026}
T.~Sj\"ostrand \etal,
\newblock JHEP{} 05~(2006)~26\relax
\relax
\bibitem{jhep:0101:010}
G.~Corcella \etal,
\newblock JHEP{} 01~(2001)~010\relax
\relax
%\bibitem{pr:d55:1280}
%H.L. Lai \etal, \relax
%\newblock Phys.\ Rev. D 55 (1997) 1280\relax
%\relax
%\bibitem{grv}
%M. Gl\"uck, G. Reya and A. Vogt,
%\newblock Phys.\ Rev.\ D~45~(1992)~3986;\relax
%\\
%M. Gl\"uck, G. Reya and A. Vogt,
%\newblock Phys.\ Rev.\ D~46~(1992) 1973\relax
%\relax
%
\end{thebibliography}
\end{document}